\documentclass[11pt,singlecolumn]{revtex4-1}
\pdfoutput=1

\usepackage{graphicx}
\usepackage{dcolumn}
\usepackage{bm}
\usepackage{color}
\pdfoutput=1

\begin{document}

\title{Enhanced valley polarization in WS$_2$/LaMnO$_3$ heterostructure}
\author{Jianchen Dang}
\author{Mingwei Yang}
\author{Xin Xie}
\author{Zhen Yang}
\author{Danjie Dai}
\author{Zhanchun Zuo}
\affiliation{Beijing National Laboratory for Condensed Matter Physics, Institute of Physics, Chinese Academy of Sciences, Beijing 100190, China}
\affiliation{CAS Center for Excellence in Topological Quantum Computation and School of Physical Sciences, University of Chinese Academy of Sciences, Beijing 100049, China}
\author{Can Wang}
\email{canwang@iphy.ac.cn}
\author{Kuijuan Jin}
\email{kjjin@iphy.ac.cn}
\author{Xiulai Xu}
\email{xlxu@iphy.ac.cn}
\affiliation{Beijing National Laboratory for Condensed Matter Physics, Institute of Physics, Chinese Academy of Sciences, Beijing 100190, China}
\affiliation{CAS Center for Excellence in Topological Quantum Computation and School of Physical Sciences, University of Chinese Academy of Sciences, Beijing 100049, China}
\affiliation{Songshan Lake Materials Laboratory, Dongguan, Guangdong 523808, China}

\begin{abstract}

Monolayer transition metal dichalcogenides have attracted great attentions for potential applications in valleytronics. However, the valley polarization degree is usually not high because of the intervalley scattering. Here, we demonstrate a largely enhanced valley polarization up to 80\% in monolayer WS$_2$ under non-resonant excitation at 4.2 K using WS$_2$/LaMnO$_3$ thin film heterostructure, which is much higher than that for monolayer WS$_2$ on SiO$_2$/Si substrate with a valley polarization of 15\%. Furthermore, the greatly enhanced valley polarization can be maintained to a high temperature of about 160 K with a valley polarization of 53\%. The temperature dependence of valley polarization is strongly correlated with the thermomagnetic curve of LaMnO$_3$, indicating an exciton-magnon coupling between WS$_2$ and LaMnO$_3$. A simple model is introduced to illustrate the underlying mechanisms. The coupling of WS$_2$ and LaMnO$_3$ is further confirmed with an observation of two interlayer excitons with opposite valley polarizations in the heterostructure, resulting from the spin-orbit coupling induced splitting of the conduction bands in monolayer transition metal dichalcogenides. Our results provide a pathway to control the valleytronic properties of transition metal dichalcogenides by means of ferromagnetic van der Waals engineering, paving a way to practical valleytronic applications.

\end{abstract}
\maketitle

\section{Introduction}

 Monolayer transition metal dichalcogenides (TMD) with a broken inversion symmetry possess two degenerate inequivalent valleys\cite{ISI:000303662500015,ISI:000295872800002}, offering a new degree of freedom of electrons, in addition to the charge and spin\cite{ISI:000306099900014,ISI:000307359600007,ISI:000307359600006,ISI:000303662500666,ISI:000303662500888}. By selectively pumping one of the two valleys with circularly polarized light, an imbalanced carrier population between +K and -K valleys can be created, leading to the valley-polarized light emission\cite{ISI:000307359600007,ISI:000307359600006}. Realizing a high valley polarization is an essential step and has been pursued many years for developing valleytronic devices\cite{ISI:000251451000055,ISI:000257289500096,ISI:000358857800010,ISI:000371946300007}. So far, the valley polarization of monolayer TMD reported is not high\cite{ISI:000340097900022,ISI:000371529600001}, which is mainly attributed to intervalley scattering process from the selectively pumped valley to the opposite one\cite{ISI:000363142100001,ISI:000311967000031}.
Several works have been reported to increase the valley polarization by using various methods, such as high magnetic field, low temperature, resonant excitation, optical or electrical doping and so on\cite{ISI:000307359600007,ISI:000307359600006,ISI:000349934700017,ISI:000348364000007,ISI:000377962500117,ISI:000377642700031,ISI:000462158000009,ISI:000389963200086}. Most methods require harsh conditions and the improvements are still not good enough. Therefore, more compact and practical approaches are desired for further applications of valleytronics.

Recently, increasing attentions have been gathered in hybrid heterostructure engineering based on TMD\cite{ISI:000428387700006,ISI:000401662800002,ISI:000454463000042,ISI:000536167500001}. By constructing van der Waals heterostructures, the properties of TMD are modulated by the additional layer or substrate, particularly the valley properties\cite{ISI:000529881300024,ISI:000378347800017,ISI:000406868800013,ISI:000485685900002,ISI:000419752300010,ISI:000435524300072}. The interactions between TMD and the adjacent materials, such as charge transfer, proximity effect, etc.,\cite{ISI:000536167500001,ISI:000406868800013,ISI:000485685900002} provide an alternative approach to controlling the valley polarizations in TMD. For example, in a monolayer TMD and graphene heterostructure, a high degree of valley polarization has been achieved due to the fast charge and energy transfer processes from monoalyer TMD to graphene\cite{ISI:000454463000042,ISI:000536167500001}. In addition, magnetic proximity effect from ferromagnetic substrates can yield an interfacial magnetic exchange field in monolayer TMD, offering an effective way for valley control\cite{ISI:000438343600080,ISI:000405387100028,ISI:000596017800003}. A giant and tunable valley splitting has been predicted theoretically and realized experimentally in monolayer TMD and ferromagnetic substrate heterostructures\cite{ISI:000406868800013,ISI:000485685900002,ISI:000369978800021}. More interestingly, a signature of interlayer exciton-magnon coupling has been observed in antiferromagnet-semiconductor van der Waals hetstructures\cite{ISI:000541691200073}. However, there are few studies reporting on improving valley polarization in monolayer TMD and ferromagnet heterostructures\cite{ISI:000419752300010}, although valley splitting control in this hybrids systems have been intensively discussed\cite{ISI:000406868800013,ISI:000485685900002,ISI:000405387100028}.

  In this work, we demonstrate a greatly enhanced valley polarization up to 80\% in monolayer WS$_2$ at 4.2 K by constructing WS$_2$/LaMnO$_3$ heterostructure. The LaMnO$_3$ (LMO) ferromagnetic film was grown by using a pulsed laser deposition (PLD) system. The polarization-resolved photoluminescence (PL) measurements were carried out for monolayer WS$_2$ both on LMO and SiO$_2$ for comparison. The enhanced valley polarization can be maintained to a high temperature, with a high value of 53\% at 160 K. More importantly, the temperature dependent valley polarization strongly follows the thermomagnetic curve of LMO, indicating an exciton-magnon coupling. Furthermore, two emerging interlayer excitons with opposite valley polarizations are observed in such hybrid heterostructure, providing a new platform for investigating valley dynamics in monolayer TMD. Our work provides a compact and effective approach to enhancing the valley polarization and to controlling the valley properties.

\section{Results and discussion}

\begin{figure}
\includegraphics[scale=0.5]{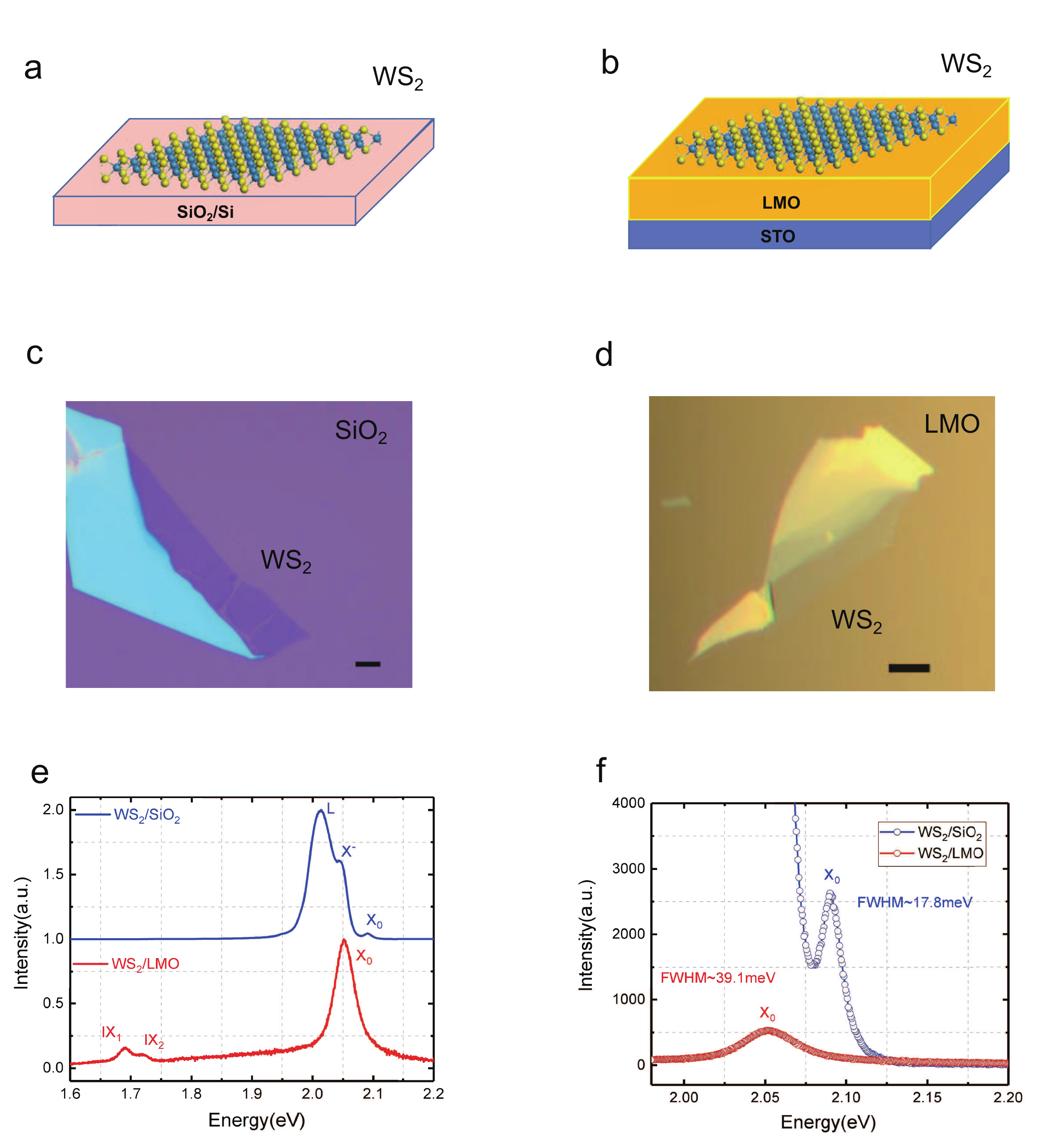}
\caption{\label{f1} Characterization of the monolayer WS$_2$ on the different substrates. (a,b) Schematic diagrams of the monolayer WS$_2$ on (a) the SiO$_2$/Si substrate and (b) the ferromagnetic LMO substrate. Optical microscopy images of the samples with a scale bar of 5 $\mu$m on (c) SiO$_2$/Si substrate and (d) LMO substrate. (e) Normalized PL spectra of monolayer WS$_2$ measured at 4.2 K with an excitation wavelength of 532 nm. Exciton (X$_0$), trion (X$^-$) and localized state(L) are observed on the SiO$_2$/Si substrate (blue line); Exciton (X$_0$) and two interlayer excitons (IX$_1$ and IX$_2$) arise on the LMO substrate (red line). The spectra are shifted for clarity. (f) Comparison of PL peaks of X$_0$ for different substrates. The linewidths (full width at half maximum, FWHM) of the X$_0$ are 17.8 meV for  WS$_2$/SiO$_2$ (blue line) and 39.1 meV for WS$_2$/LMO (red line), respectively.}
\end{figure}

Figure 1a,b show schematic diagrams of the monolayer WS$_2$ on SiO$_2$/Si and the ferromagnetic LMO substrates, respectively. After the deposition with PLD, the structure and the thickness of the LMO film are characterized with X-ray diffraction, as shown in Supplementary Information (SI) Fig. S1. The monolayer WS$_2$ are mechanically exfoliated from bulk and then transferred onto the LMO and SiO$_2$/Si substrates. The optical microscopy images of the samples are shown in Figs. 1c and d, and the monolayer thickness is also confirmed with the atomic force microscope measurement, as shown in SI Fig. S2. Figure 1e shows the normalized micro-PL spectra of monolayer WS$_2$ on SiO$_2$ and LMO substrates under a non-resonant excitation with a 532-nm continuous laser at 4.2 K. Three distinct peaks at 2.089 eV, 2.045 eV and 2.013 eV arise in WS$_2$/SiO$_2$, which can be assigned to exciton (X$_0$), trion (X$^-$) and localized state (L)\cite{ISI:000348619000070,ISI:000360382300003,ISI:000419226900007}, respectively. In contrast, there is only one peak at 2.052 eV presented in WS$_2$/LMO. This peak originates from X$_0$ emission, which is confirmed by the reflectance contrast spectra, as shown in SI Fig. S3. Compared to WS$_2$ on SiO$_2$ substrate, the peak of WS$_2$ on LMO exhibits redshift and broadening in both PL and reflectance spectra. The observed red shift is expected from dielectric screening of LMO substrate, which screens the electric field between electrons and holes and leads to a strong reduction in the bandgap but a modest reduction of the exciton binding energy of the monolayer WS$_2$\cite{WOS:000341115700020,ISI:000400561500001}. The absence of X$^-$ and localized state, as well as a 5-fold suppression of the X$_0$ intensity comparing with WS$_2$/SiO$_2$ are observed as shown in Fig. 1e and f, which result from the additional recombination processes such as charge transfer\cite{ISI:000536167500001,ISI:000419226900007,ISI:000381959100067}. The linewidth on WS$_2$/LMO with 39.1 meV is much broader than 17.8 meV on WS$_2$/SiO$_2$, which may result from the enhanced thermal phonon scattering, impurity and charge transfer\cite{WOS:000359613700028}. Among these factors, the charge transfer may be the dominant factor for the broadening, as discussed in more detail in SI.

\begin{figure}
\includegraphics[scale=0.5]{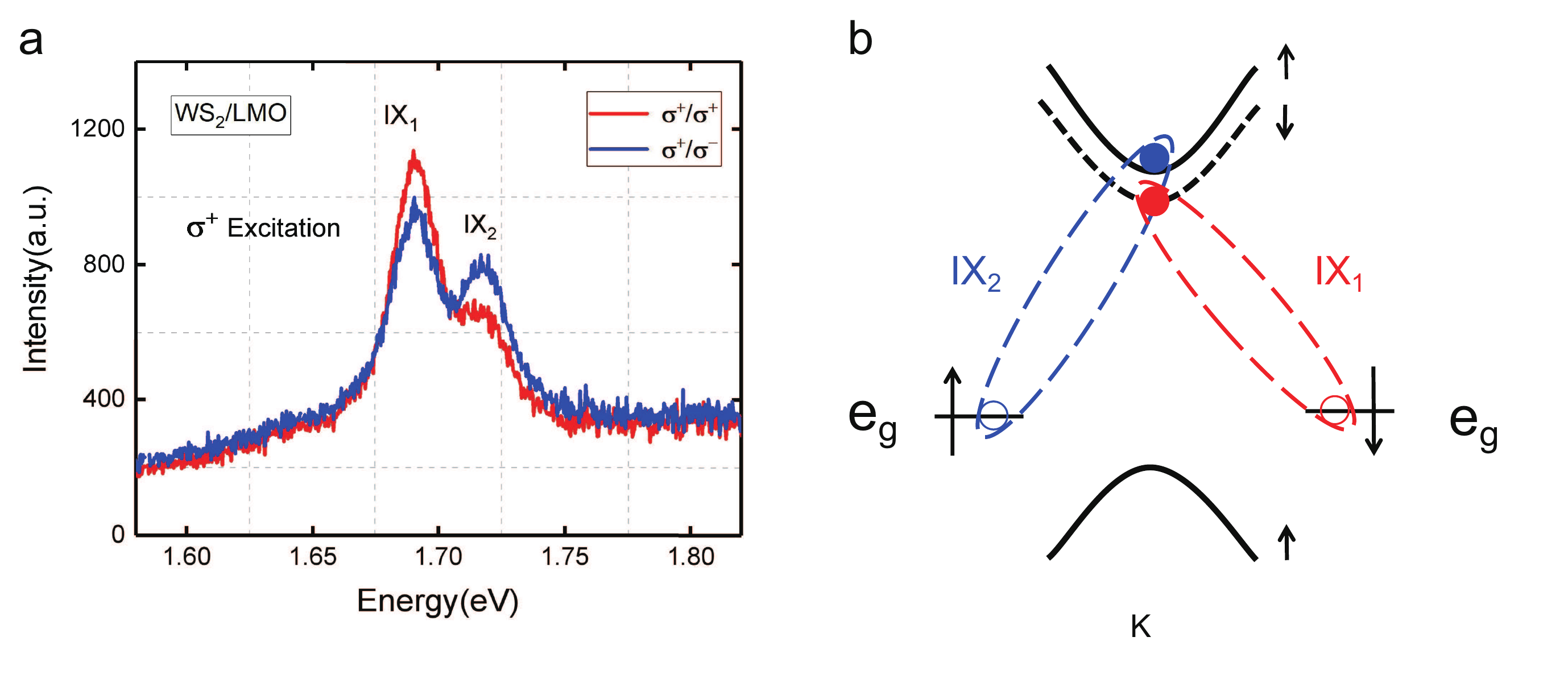}
\caption{\label{f2} PL spectra of double interlayer excitons and Schematic diagram of recombination mechanisms. (a) Polarization-resolved PL spectra of double interlayer excitons in monolayer WS$_2$/LMO heterostructure. (b) Schematic of energy level of the double interlayer excitons in monolayer WS$_2$/LMO. A small spin-orbit splitting arises in the conduction band of monolayer WS$_2$, and e$_g$ represents the  energy level of LMO. The arrow directions indicate the spin-up and spin-down states. The red solid (empty) circle represents the electron (hole) of interlayer exciton IX$_1$ and the blue represents the interlayer exciton IX$_2$. The dashed ellipses indicate the formation of the double interlayer excitons with opposite valley polarizations.}
\end{figure}

In addition, in the lower-energy range compared to intrinsic excitons, two peaks arised at 1.689 eV and 1.718 eV on WS$_2$/LMO heterostructure. The energy separation between the two peaks is about 29 meV, which is in agreement with the conduction band splitting of WS$_2$ caused by spin-orbit coupling\cite{ISI:000303662500015,ISI:000323573600010}. So the two peaks may originate from the interlayer excitons and the splitting can be attributed to the interlayer fine structures\cite{ISI:000507151600090,ISI:000433404500070,ISI:000456417000003}. To further verify the origin of the double peaks in WS$_2$/LMO heterostructure, the polarization-resolved PL signals are non-resonantly excited by a $\sigma^+$ polarized laser at 532 nm (2.33 eV) and analyzed with configurations of co-polarized $\sigma^+$ and cross-polarized $\sigma^-$ luminescence. Figure 2a shows the polarization-resolved PL spectra. The double peaks exhibit opposite valley polarizations, which is very similar to the behaviour of double interlayer excitons in TMD heterobilayers\cite{ISI:000507151600090,ISI:000433404500070,ISI:000456652900019,ISI:000474369000006,ISI:000411043500008,ISI:000526408800050}. In WS$_2$/LMO heterostructure, the splitting of conduction band leads to the formation of the two interlayer excitons with opposite valley polarizations, as indicated by the blue and red dashed ellipses in Fig. 2b. Moreover, photoluminescence excitation measurement is performed to support the assignment of PL emission to the interlayer exciton, as shown in SI Fig. S4. Therefore, the double peaks with opposite valley polarizations originate from the interlayer exciton between LMO and monolayer WS$_2$. Besides, the double interlayer excitons in such TMD/ferromagnet heterostructures also verifies the existence of charge transfer from WS$_2$ to LMO substrate, which might enhance the valley polarization of exciton in monolayer WS$_2$\cite{ISI:000454463000042,ISI:000536167500001}.

\begin{figure}
\includegraphics[scale=0.5]{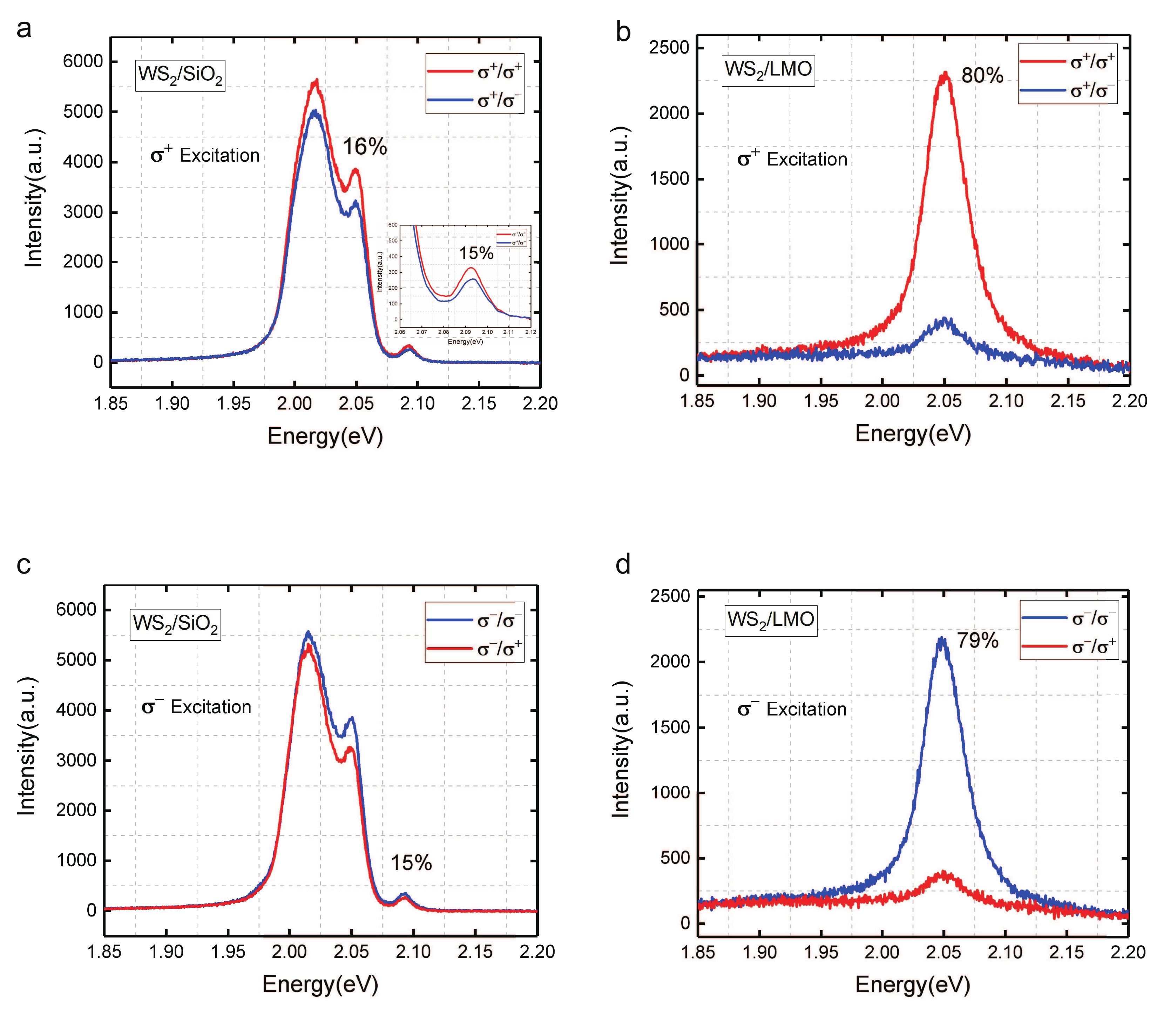}
\caption{\label{f3} Polarization-resolved PL spectra of the monolayer WS$_2$ on the different substrates. Polarization-resolved PL spectra of monolayer WS$_2$ on (a) SiO$_2$/Si and (b) LMO substrates under $\sigma^+$ excitation. The polarization degree of X$_0$ for SiO$_2$/Si and LMO substrates are 15\% and 80\%, respectively. The polarization degree of X$^-$ for SiO$_2$/Si is about 16\%. Polarization-resolved PL spectra of monolayer WS$_2$ on (c) SiO$_2$/Si and (d) LMO substrates under $\sigma^-$ excitation. The polarization degrees of X$_0$ for SiO$_2$/Si and LMO substrates are 15\% and 79\%, respectively. Red lines represent the $\sigma^+$ light emission and blue lines represent the $\sigma^-$ light emission.}
\end{figure}

Figure 3 shows the polarization-resolved PL spectra in monolayer WS$_2$/SiO$_2$ and WS$_2$/LMO. Based on the optical selection rules in monolayer TMD, the valley polarization degree P$_c$ can be depicted by\cite{ISI:000307359600007,ISI:000462158000009},
\begin{equation}
 P_c =\frac{I (\sigma^+)- I (\sigma^-)}{I (\sigma^+)+ I (\sigma^-)}
\label{wkb}
\end{equation}
where I ($\sigma^+$) and I ($\sigma^-$) are the $\sigma^+$ and $\sigma^-$ PL intensities. Here, we only focus on X$_0$ emission. The valley polarization degree of X$_0$ in WS$_2$/SiO$_2$ is about 15\%, which is extracted from the polarization-resolved PL spectra in Fig. 3a with $\sigma^+$ excitation according to Eq. 1, similar to previous results\cite{ISI:000340097900022,ISI:000368379000001}. In contrast, the valley polarization in WS$_2$/LMO is largely enhanced, with a high value up to about 80\%, as shown in Fig. 3b. The similar results are obtained under $\sigma^-$ excitation, shown in Figs. 3c and d. Note that no distinguishable valley splitting is observed in our WS$_2$/LMO heterostructure.

To clarify the underlying mechanisms for the greatly enhanced valley polarization, temperature dependent valley polarization degrees were measured for both two samples, as shown in Fig. 4a. Both samples exhibit a monotonic increase of the valley polarization with decreasing temperature. For WS$_2$/LMO, the valley polarization increases drastically from 200 K to 100 K, and then maintains a high value of about 80\%. In contrast, for WS$_2$/SiO$_2$, the valley polarization possesses low values, which is only 15\% at 4.2 K. To determine the origin of the enhanced valley polarization, the temperature-dependent magnetization of LMO substrate was also measured using a vibrating sample magnetometer of physical properties measurement system, as indicated by the blue line in Fig. 4a. Remarkably, the temperature dependent valley polarization of WS$_2$/LMO is strongly correlated with the thermomagnetic curve of LMO when superimposed on each other. It indicates that the enhanced valley polarization in monolayer WS$_2$ most likely originates from the magnetic proximity effect from LMO substrate. At higher temperatures there is a slight deviation between the two curves.

\begin{figure}
\includegraphics[scale=0.5]{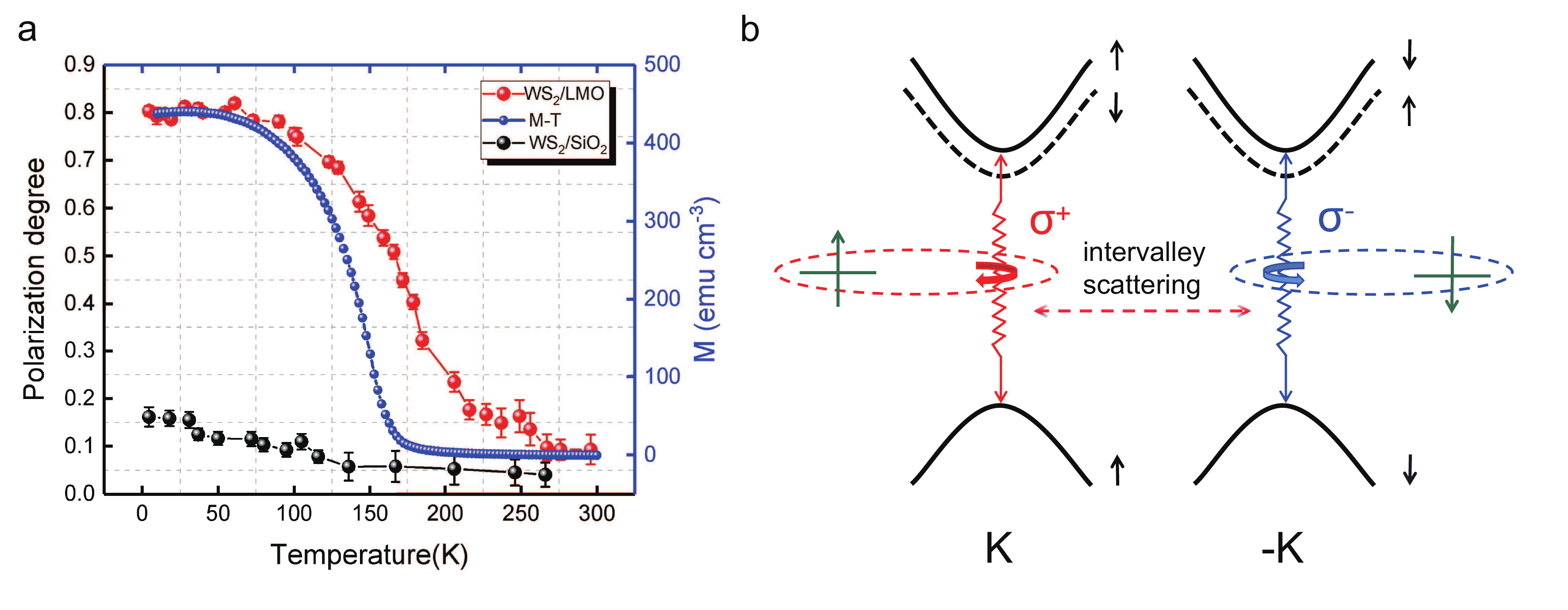}
\caption{\label{f4} Temperature dependent valley polarization of the monolayer WS$_2$ on the different substrates and a simplified model for the enhanced valley polarization. (a) Temperature dependent valley polarization of X$_0$ in WS$_2$ on LMO substrate (red solid circles), SiO$_2$/Si substrate (black solid circles) and temperature dependent magnetization (M-T) of LMO (blue solid circles) superimposed on each other. The left and right vertical axises represent the polarization degree and magnetization (M), respectively. (b) Schematic diagram of transitions in monolayer WS$_2$/LMO, containing the optical transitions obeying optical selection rules, intervalley scattering process and the coupling between exciton and magnon. The dashed lines in the conduction band depict spin-forbidden state for the ground state of monolayer WS$_2$. The black arrows represent the spin states in monolayer WS$_2$ and the green represent the magnon (quanta of spin wave) in LMO ferromagnetic substrate. The dashed ellipses denote the coupling between exciton and magnon.}

\end{figure}

Next, we will discuss the reasons for the enhanced valley polarization in WS$_2$/LMO heterostructure. Two mechanisms might be involved. When a $\sigma^+$ ($\sigma^-$) laser selectively excites the exciton transitions in the +K (-K) valley, the corresponding recombination emission should carry single handedness\cite{ISI:000307359600007,ISI:000307359600006}, namely the unity valley polarization, because of the optical selection rule. However, as schematically shown in Fig. 4b, when the +K valley is selectively pumped with $\sigma^+$ excitation, the generated PL light emissions contain not only $\sigma^+$ component from exciton transitions in +K valley but also $\sigma^-$ component due to intervalley scattering of excitons from +K to -K valley, generating a non-unity valley polarization\cite{ISI:000363142100001,ISI:000311967000031}. When the monolayer WS$_2$ is transferred on the LMO substrate, the charge transfer occurs, which is demonstrated by the PL intensity suppression, linewidth broadening (Fig. 1f) and the formation of interlayer excitons (Fig. 1e and Fig. 2). The charge transfer process provides new decay channels for WS$_2$ excitons. After the generation process of excitons in +K valley pumped by $\sigma^+$ light, the exciton populations will now rapidly recombine and transfer the photoexcited energy to LMO substrate non-radiatively or be used to form interlayer excitons instead of being scattered to -K valley. Therefore, the proportion of intervalley scattering is reduced, leading to the increase of valley polarization. 

On the other hand, interlayer exciton-magnon coupling might be another dominant factor responsible for the enhanced valley polarization. Magnon is defined as quanta of spin wave\cite{gong2019two}. The exciton-magnon coupling is essentially one of the mechanisms of the magnetic proximity effect, which can occur at the heterointerface formed by an atomically-thin semiconductor and magnetic materials. For our samples, the temperature dependent valley polarization strongly follows the thermomagnetic curve of LMO. We suggest that it is caused by exciton-magnon coupling, which is related to the locking of valley excitons with different polarizations and magnons with different spin orientations. More specifically, the K valley excitons are locked with the spin-up magnons, and the -K valley excitons are locked with the spin-down magnons, as shown in Fig. 4b. Below the Curie temperature, the spin orientation is fixed and such locking between exciton and magnon will strongly suppress intervalley scattering of exciton, leading to an enhanced valley polarization. Note that the exciton-magnon coupling in LMO substrate might be related to the band structure and lattice match to the WS$_2$, the specific termination surface, crystal orientation and the stoichiometry of LMO film as discussed in SI. With the increase of temperature, the spin tends to be chaotic and the coupling of exciton-magnon will weaken. Therefore, the suppression of intervalley scattering due to the exciton-magnon coupling becomes weak, leading to a decrease of the valley polarization. As the temperature further increases, the former mechanism (charge transfer process) dominates, causing a slight deviation between the temperature dependent valley polarization and the thermomagnetic curve of LMO, as shown in Fig. 4a. Similar results are observed in three other monolayer WS$_2$ samples with random angles on the same LMO, as shown in SI Fig. S5, suggesting the exciton-magnon coupling is not dependent on the rotation alignment between monolayer WS$_2$ and LMO due to the in-plane magnetic isotropy of LMO (001) film (shown in SI Fig. S6). For the four samples, there exists a little discrepancy in the valley polarization degree, which may result from the quality of the samples or the different heterointerface that occur during the transfer process, thereby affecting the valley polarization degree. Additionally, we also measured the valley polarization degree of monolayer MoS$_2$, WSe$_2$ and MoSe$_2$ on LMO and SiO$_2$ substrates. Neither obvious increase of valley polarization degree nor interlayer excitons was observed in the three materials with LMO substrate, where the reason need to be investigated further.

\begin{figure}
\includegraphics[scale=0.8]{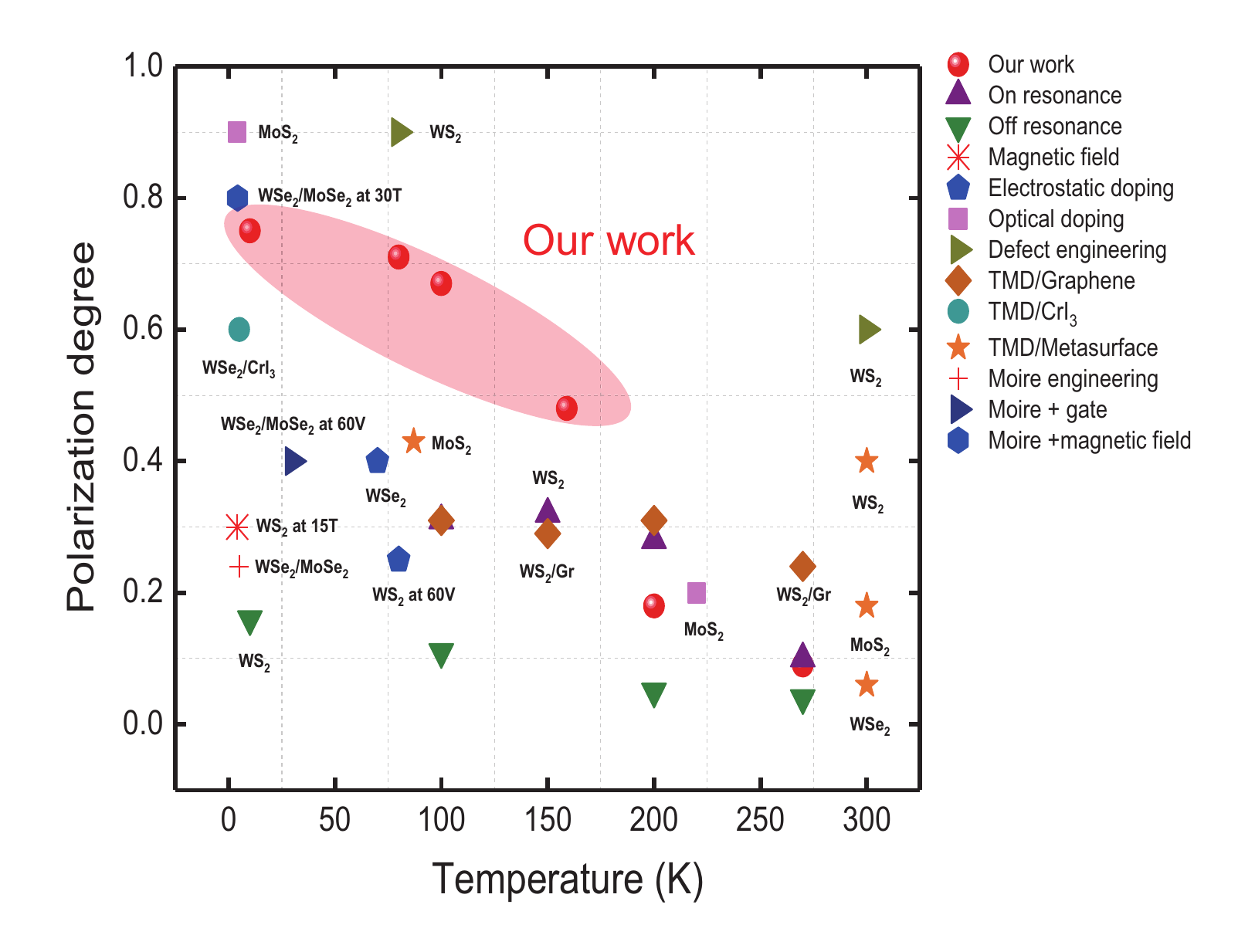}
\caption{\label{f5} Comparison on the valley polarization of TMD materials. solid circles denote the average valley polarization degree of the four samples from our work and the rest of dots with other colors and shapes represent the previous reported values, in which different methods have been explored, such as on/off resonance excitation from Ref.[12], applied magnetic field from Ref.[21], electrostatic doping from Refs.[20] and [58], optical doping from Refs.[4] and [55], defect engineering from Ref.[57], heterostructures (TMD/Gr from Ref.[25], TMD/CrI$_3$ from Ref.[30], and moire patterns from Ref.[48]), moire + gate from Ref.[61], moire + magnetic field from Ref.[54], and TMD/metasurface from Refs.[62], [64] and [65].}

\end{figure}

In the following, we compare our work with the previous results on valley polarizations in monolayer WS$_2$ and other TMD materials. Up to now, many methods such as strong magnetic field, resonant excitation, heterostructure engineering and defect and strain engineering have been applied to improve valley polarization, as shown in Fig. 5. With applied out-of-plane magnetic field, the polarization degree exhibits monotonically increasing and has been optimized to 30\% under 15 T at 4 K in monolayer WS$_2$\cite{ISI:000389963200086}. However, this method requires high magnetic field, and the enhancement of polarization degree is not very high.  Under resonant excitation, the valley polarization of 40\% in monolayer  WS$_2$ at 10 K has been reported, while the valley polarization under off-resonant excitation is only about 16\%\cite{ISI:000340097900022}. Additionally, a near-unity valley polarization has been reported in monolayer MoS$_2$ with resonant excitation\cite{ISI:000307359600007,ISI:000307273100002}. But with different substrates such as hBN or SiO$_2$ with different thicknesses, the valley polarizations in monolayer MoS$_2$ show big difference which may result from the different influences of substrate materials on the optical properties of monolayer MoS$_2$\cite{ISI:000306099900014,ISI:000307359600007,ISI:000307359600006,ISI:000307273100002}. Despite of the difference in different materials, the resonant excitation requires sophisticated measurement setup to filter out the pumping light. By all means, the reported valley polarizations under non-resonant excitation are very low. By heterostructure engineering, the polarization degree of 41\% has been reported in WS$_2$/Graphene (WS$_2$/Gr) at 4 K, and 25\% has been reported at room temperature with resonant excitation in WS$_2$/Gr\cite{ISI:000536167500001}. However, these heterostuctures are difficult to fabricate and the optimization is not obviously improved. Defect and strain engineering can also tune valley polarization, and 90\% valley polarization at 80 K has been reported in CVD-grown monolayer WS$_2$ single crystals riched in defects\cite{ISI:000518024700009}. However, the defect and strain engineering are difficult to precisely control and may degrade the valley polarization\cite{ISI:000408520900047}. In addition, electrostatic doping, moire patterns, and plasmonic architectures based on TMD have also be used to produce and tune valley polarization, as shown in Fig. 5\cite{ISI:000306099900014,ISI:000307359600007,ISI:000307359600006,ISI:000462158000009,ISI:000419752300010,ISI:000433404500070,ISI:000456652900019,ISI:000478617900025,ISI:000425594600004,ISI:000442206400013,ISI:000369810000035,ISI:000490638700001,ISI:000460426900043,ISI:000415323000024,ISI:000459333300014,ISI:000430642500017}.
But all these methods either require harsh conditions, such as high magnetic field, sophisticated measurement setup and difficult fabrication processes, or exhibit low-efficiency optimization of valley polarization. Thus an efficient and convenient method is more important and practical on manipulating the valley polarization. By utilizing monolayer WS$_2$/LMO heterostructure here, the valley polarization exhibits a high value up to 80\% under non-resonant excitation at 4.2 K, and keeps at a high value of 53\% even when the temperature is increased to about 160 K. Therefore, we believe our method provides an efficient and convenient way to manipulate the valley polarization, making the valleytronic devices more practical.

\section{Conclusion}

In summary, we experimentally demonstrated the greatly enhanced valley polarization up to 80\% in monolayer WS$_2$/LMO heterostructure at 4.2 K with off-resonance excitation by constructing WS$_2$/LMO heterostructure. And the enhanced valley polarization can be maintained to a higher temperature, up to 160 K (about 53\%). Furthermore, the temperature-dependent valley polarization highly follows the thermomagnetic curve of LMO, indicating the exciton-magnon coupling. A simplified model is introduced here, in which two mechanisms are proposed to be responsible for the enhanced valley polarization. In addition, double interlayer excitons with opposite valley polarizations in such hybrid heterostructure were observed, providing a new platform for valleytronics and spintronics. Further studies will be required to fully understand the underlying mechanisms. Nevertheless, we have shown a strong evidence experimentally that utilizing WS$_2$/LMO heterostructure engineering, the valley polarization can be greatly improved with non-resonant excitation, which provides an effective and convenient approach for valley control, promoting the development of the applications of monolayer TMD valleytronics. Moreover, it can be predicted that the intriguing phenomena will motivate more studies on the heterostructure formed by the ferromagnetic materials and monolayer TMD.

\begin{acknowledgments}
This work was supported by the National Key Research and Development Program of China (Grant No. 2021YFA1400700), the National Natural Science Foundation of China (Grants Nos. 62025507, 11934019, 11721404, 11874419, 61775232, 11874412, and 12174437), the Key R$\&$D Program of Guangdong Province (Grant No. 2018B030329001), the Strategic Priority Research Program (Grant No. XDB28000000) of the Chinese Academy of Sciences.
\end{acknowledgments}

\end{document}